\documentclass{aa}
\usepackage{psfig}
\usepackage{graphicx}

\def\brona{1RXS J171824.2--402934}
\def\bronb{1RXS J170854.4--321857}
\def\ecs{erg~cm$^{-2}$s$^{-1}$}
\def\lum{erg~s$^{-1}$}

\begin{document}

\title{On the nature of two low-$\dot{M}$ X-ray bursters: \\ \bronb\ and \brona}

\titlerunning{On the nature of two low-$\dot{M}$ X-ray bursters} 
\authorrunning{J.J.M. in 't Zand et al.}

\author{
J.J.M.~in~'t~Zand\inst{1,2},
R. Cornelisse\inst{3}
\& M.~M\'{e}ndez\inst{1,2}
}


\institute{     SRON National Institute for Space Research, Sorbonnelaan 2,
                NL - 3584 CA Utrecht, the Netherlands 
	 \and
                Astronomical Institute, Utrecht University, P.O. Box 80000,
                NL - 3508 TA Utrecht, the Netherlands
         \and
                Department of Physics and Astronomy, University of Southampton,
                Hampshire SO17 1BJ, U.K.
	}

\date{Draft; version \today}

\abstract{We carried out Chandra observations of two low-luminosity
low-mass X-ray binaries, \bronb\ and \brona, for which previously
single X-ray bursts had been detected with the Wide Field Cameras
(WFCs) on board BeppoSAX. Both were detected in our Chandra
observations in an actively accreting state three to eight years after
the X-ray bursts, with 0.5--10 keV luminosities between
$5\times10^{34}$ and $2\times10^{36}$~\lum. The apparently persistent
nature is remarkable for \brona\ given its low luminosity of
10$^{-3}$~$L_{\rm Edd}$. The persistence of both sources also
distinguishes them from 5 other low-$L$ bursters,
which have also been seen during bursts with the WFCs but were not
detected during Chandra observations above a luminosity of
10$^{33}$~\lum. Those are probably transient rather than persistent
sources. \keywords{X-rays: binaries -- X-rays: bursts -- X-rays:
individual: \bronb = 4U~1705--32 = 1M 1704--321 = 1H 1659--317 ;
\brona = RX J1718.4--4029}}

\maketitle 

\section{Introduction}
\label{intro}

Currently almost 200 low-mass X-ray binaries (LMXBs) are known in our
galaxy (150 are cataloged by Liu et al. 2001 and several tens have
been discovered since then). Roughly half exhibit type-I X-ray bursts
(e.g., In 't Zand et al. 2004) which are due to thermonuclear flashes
on neutron star surfaces (for reviews, see Lewin et al. 1993, Bildsten
1998 and Strohmayer \& Bildsten 2005).  Most (type-I) X-ray bursters
have accretion rates in excess of 1\% of the Eddington limit for a
canonical neutron star ($10^{-8}~M_\odot$yr$^{-1}$) and burst
recurrence times between hours and months. However, at least seven
bursters may have lower accretion rates (see Tables~\ref{tab1} and
\ref{tab2}). These were discovered through observations with the Wide
Field Cameras (WFCs; Jager et al. 1997) on BeppoSAX (Boella et
al. 1997). During its 1996--2002 operational life this instrument
gathered 1 to 12 Msec (depending on exact position) of exposure along
the Galactic plane which resulted in the detection of rare bursters
like the seven bursters without persistent emission (In 't Zand et
al. 1998, Kaptein et al. 2000, Cocchi et al. 2001, Cornelisse et
al. 2002a and In 't Zand et al. 2004). Each of these bursters
exhibited a single burst. The flux limit on the persistent emission of
these rare bursters translates (for a canonical 8~kpc distance) to a
luminosity limit of $\sim10^{36}$~\lum\ or roughly 1\% of the
Eddington limit. Five bursters were previously unknown sources while
two, \bronb\ and \brona, were detected with ROSAT (Kaptein et
al. 2000; In 't Zand et al. 2004).

Cornelisse et al. (2002b) followed up four of the five low-$L$
bursters without ROSAT counterparts, with the Chandra observatory for
5 ksec each. No unique counterparts could be identified.  In three
cases multiple candidate counterparts were identified with persistent
luminosity levels of $10^{32-33}$~\lum\ for 8~kpc distances, while in
the fourth case a similar-valued upper limit was inferred. These
levels are typical for neutron star transients in quiescence (Verbunt
et al. 1994; Asai et al. 1998; Campana et al. 1998; Rutledge et
al. 1999; Menou et al. 1999) suggesting that, if they are truly
type-I X-ray bursters, they are transient sources whose peak flux
during outburst was too low to be detected with monitoring devices
such as the WFCs or the All-Sky Monitor (ASM) on RXTE (Levine et
al. 1996).

The two low-$L$ bursters with ROSAT counterparts may be of a different
nature. Since ROSAT detected them several years prior to the WFCs, the
suggestion is there that these are {\em persistent} low-luminosity
systems. In order to confirm this, we followed them up with Chandra
with the Advanced CCD Imaging Spectrometer spectroscopic array
(ACIS-S) in the focal plane. Also, we carried out a more detailed
analysis of the WFC-detected burst of \bronb, searched through RXTE
data for detections, and browsed the literature for historical
detections. The results, as well as a discussion of the implication of
these observations for the nature of these sources, are presented
here.

\section{Observations of \bronb}
\label{bronb}

\subsection{BeppoSAX/WFC}
\label{bronbburst}

\begin{figure}[t]
\includegraphics[width=\columnwidth,angle=0]{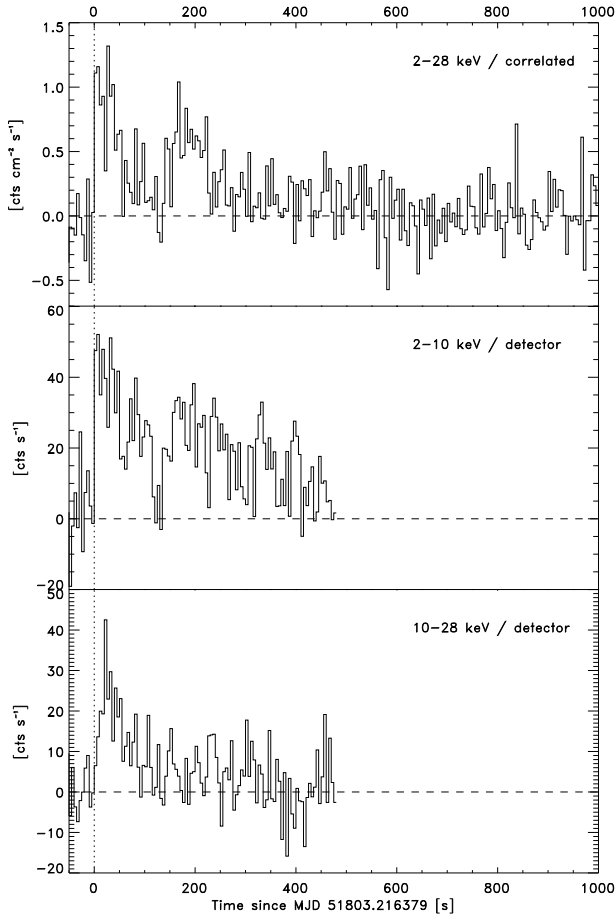}
\caption[]{WFC light curves of \bronb. {\it Top:} in total
bandpass, from reconstructed source images ('correlated' with coded
aperture; see Jager et al. 1997).  {\it Middle and bottom:} at low and
high photon energies, from photon rates on the detector. Detector
photon rates are not calibrated but we have chosen to present those
rather than the imaged rates as in the top panel, because they have a
slightly better statistical quality which for this faint burst enables
easier visual comparison of the profiles in two bandpasses. These
profiles are cut at $t=500$~s because the subsequent data show a burst
from a different source in the same field of view.  The time
resolution is 5~s.
\label{rxj1708proflc}}
\end{figure}

\begin{figure}[t]
\includegraphics[width=\columnwidth,angle=0]{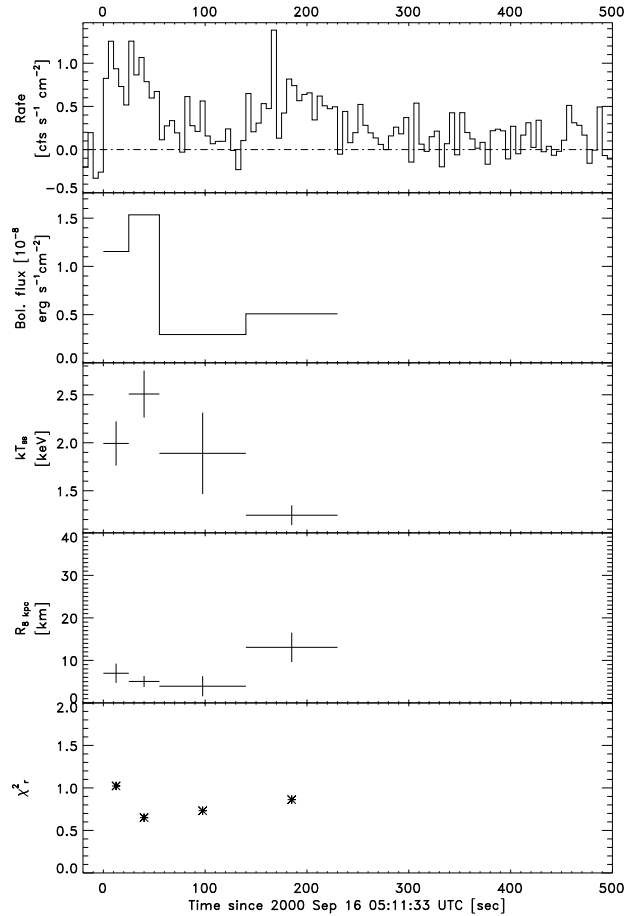}
\caption[]{Time-resolved spectroscopy of the burst from \bronb. 
\label{rxj1708prof}}
\end{figure}

The burst from \bronb\ started at 05:11:35 UT on Sep. 16, 2000 (MJD
51803.216379) and lasted fairly long, with a duration of approximately
10 min. This burst was previously reported by In 't Zand et
al. (2004), but we here provide a more detailed
analysis. Figures~\ref{rxj1708proflc} and \ref{rxj1708prof} present
the light curve and time-resolved spectroscopy. The burst is
relatively faint. Only 4 meaningful time bins could be employed for
spectroscopy. Nevertheless, the data are consistent with a blackbody
spectrum and exhibit softening during decay. Combined with a
sub-second rise this provides ample evidence for identification as a
type-I X-ray burst (Hoffman et al. 1978). There are two peculiarities
about the burst. First, the flux shows a slower increase in 10--28 keV
than in 2--10 keV. This suggests that there is mild photospheric
radius expansion due to near-Eddington luminosities.  Unfortunately,
the faintness of the burst precludes a spectral confirmation of such
an expansion. If we translate the peak bolometric flux to the
Eddington limit of a canonical neutron star ($2\times10^{38}$~\lum\
for a hydrogen-rich and $3.8\times10^{38}$~\lum\ for a hydrogen-poor
photosphere; e.g. Kuulkers et al. 2003), we estimate a distance
between 11 and 15 kpc. Therefore, the faintness of the burst seems to
be due to a large distance. The second peculiarity involves a drop in
flux around 100~s after the onset. The indication is that the drop is
to zero. This may or may not be related to temporarily increased
absorption by intervening material. This is impossible to test with
these data.  The burst bolometric fluence is
$(1.5\pm0.5)\times10^{-6}$~erg~cm$^{-2}$ which for the distance of
13$\pm2$~kpc translates to a radiative energy output of
$3.0_{-1.5}^{+2.4}\times10^{40}$~erg.
 
We evaluated the position of the source that produced this burst from
a 500-s data stretch in the 2--10 keV bandpass and find
$\alpha_{2000.0}=17^{\rm h}08^{\rm m}52\fs5, \delta_{2000.0}=-32^\circ
19\arcmin26\arcsec$ with a 99\%-confidence error circle of radius
1\farcm8. This is 0\farcm6 from the centroid position of \bronb, well
within the WFC error circle.  The nearest other ROSAT source is
14\farcm7 from this position, which implies that source confusion is
not an issue and that the burst can be fairly confidently associated
with this source.

\subsection{Chandra/ACIS-S}
\label{bronbchandra}

\bronb\ was observed with Chandra for 13.7 ksec (net exposure time)
with ACIS-S (Garmire et al. 2003) in the focal plane starting Jan. 28,
2004, at 00:13:30 UT (obsid 4549). The pointing direction was nominal
so that the source was focused on chip S3.  The CCD frame time was set
to the nominal 3.2 s. The image shows a heavily piled-up source with
the typical hole in the center of the point spread function and
readout trails.  We determined the average of the photon positions
within a circle with a 12\farcs5 radius. This translates to
$\alpha_{2000.0}=17^{\rm h}08^{\rm m}54\fs269$,
$\delta_{2000.0}=-32^\circ 19\arcmin57\farcs13$ ($l^{II}=-7\fdg22$,
$b^{II}=+4\fdg67$) with a nominal uncertainty of 0\farcs6 (all
forthcoming uncertainties in this paper are for a confidence level of
90\% for a single parameter of interest). This is 1\farcs7 from the
ROSAT position which is consistent given that the ROSAT position has
an accuracy of 8\arcsec. Searches in red and infrared plates of the
Digital Sky Survey and in the USNO-B1.0 catalog have not revealed an
optical counterpart. The nearest star is 1\farcs6 from the centroid.
Typical magnitude limits of these searches are 20 mag. The nearest
near-infrared object in the 2MASS catalog is at 10\farcs7.

Since the pile up is heavy, we resorted to the photons in the readout
trails for a spectral analysis. The photons from the readout trails
are not subject to pile up as long as the count rate is below a few
thousand s$^{-1}$, because during CCD readout the pixels have a
readout time of only 40~$\mu$s instead of 3.2~s. We extracted the
readout source spectrum through CIAO tool {\tt acisreadcorr} with an
extraction region that is 2\farcs5 wide, starts at 25\arcsec\ from the
source centroid and ends at the CCD edges. The background spectrum was
extracted in a similar region 100 pixels to the north.  The source
spectrum contains 1161 photons and the background spectrum 243
background photons. The spectrum was rebinned so that each photon
energy bin contains at least 15 photons to ensure applicability of the
$\chi^2$ statistic. The effective exposure time is 1.16\% of the total
exposure time. An absorbed power law model fits this spectrum
satisfactorily ($\chi^2_\nu=1.24$ for $\nu=53$) and yields $N_{\rm
H}=(4.0\pm1.0)\times10^{21}$~cm$^{-2}$, a photon index of
$\Gamma=1.9\pm0.2$ and a 0.5--10 keV flux of
$(7.3\pm0.6)\times10^{-11}$~\ecs.  For a distance of 11--15 kpc, this
translates to a luminosity of $(1-2)\times10^{36}$~\lum.

The spectrum is typical of low-$L$ LMXBs. Wilson et al. (2003)
performed a Chandra study of 8 low-$L$ LMXBs, also using ACIS-S. Half
of these are X-ray bursters. The spectra of those could be modeled
with an absorbed power law with a photon index of 1.8--2.2. Such
indices are typical of faint non-quiescent neutron-star LMXBs in
general (cf., Barret, McClintock \& Grindlay 1996).

We inspected the light curve for photons in an annulus around the PSF
centroid with an inner and outer radius of 1\farcs5 and 5\farcs0
respectively. There are no noticable features at a time resolution of
32 to 320~s. We also inspected the lightcurve of the trailed photons
which allow studying frequencies up to 10$^4$~Hz due to the fast
readout. No signal was found in the power spectrum above a 3$\sigma$
upper limit in the pulsed fraction of $\sim$25\%.

\subsection{Other}

Interestingly, \bronb\ appears to have a long X-ray history prior to
ROSAT. A source was detected at the same position at 5--25~mCrab (2--6
keV) already in 1971--1973 by UHURU (4U 1705--32; Forman et al. 1978),
at $6\pm2$~mCrab (3--10 keV) with OSO-7 in 1971--1973 (1M 1704--321;
Markert et al. 1979) and at 1.6 mCrab with HEAO-1/A-1 in late 1977 (1H
1659--317; Wood et al. 1984).

\begin{figure}[t]
\includegraphics[width=\columnwidth,angle=0]{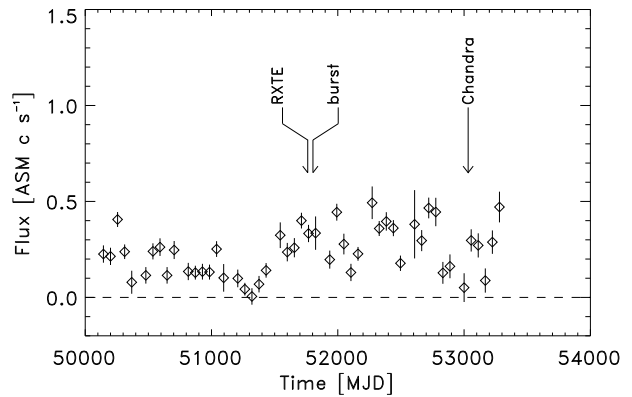}
\caption[]{ASM light curve of \bronb\ with a time resolution of 8
weeks. All data point with errors in excess of 0.1~ASM c~s${-1}$ have
been eliminated. No bias level was subtracted. The arrows indicate the
times of the WFC-detected burst, and the Chandra and RXTE pointed
observations.
\label{j1708asm}}
\end{figure}

Figure~\ref{j1708asm} shows the 3--12 keV light curve as measured
since 1996 with the ASM at a time resolution of 8 weeks, based on 2.9
Msec worth of data. There is a small signal with a factor of 10
variability which appears to change in strength at MJD 51300
(approximately one year before the WFC detected the burst). Before
that the average flux is $2.4\pm0.1$~mCrab and afterwards
$3.4\pm0.1$~mCrab (3--12 keV). For a Crab-like spectrum, as measured
with Chandra, these numbers translate to 0.5-10 keV fluxes of 7 and
$10\times10^{-11}$~\ecs. The fractional rms does not change. We
note that these flux levels are rather low for the ASM and may
be liable to systematic errors. At higher fluxes this error is
negligible. A check of the dwell data reveals that the ASM for certain
did not detect any X-ray bursts like seen with the WFC.

RXTE also made a pointed observation of \bronb. It was a
target-of-opportunity observation on Aug 6, 2000, at 02:37:20 UT for
3.4 ks after it was detected during a slew maneuver. This is 41~d
prior to the WFC-detected burst. All PCUs of the Proportional Counter
Array (PCA; Jahoda et al. 1996) were active during this observation.
No burst was detected. The overall 3--20~keV spectrum, as distributed
through the 'standard products', could be satisfactory modeled
($\chi^2_\nu=1.44$ for $\nu=37$) with an absorbed power law with
$N_{\rm H}=(1.5\pm0.7)\times10^{22}$~cm$^{-2}$ and
$\Gamma=2.09\pm0.05$.  The value for $N_{\rm H}$ is inconsistent with
that determined through the Chandra spectrum which is more sensitive
to $N_{\rm H}$ because of its lower energy coverage. If we fix $N_{\rm
H}$ to the Chandra value, the fit is only slightly worse with
$\chi^2_\nu=1.56$ ($\nu=38$) and $\Gamma=2.02\pm0.03$. The 3--20~keV
flux is $1.0\times10^{-10}$~\ecs. The unabsorbed 0.5--10 keV flux is
$1.7\times10^{-10}$~\ecs\ which is 2 times higher than during the
Chandra observation. The extrapolated 0.5--40~keV flux is
$2.4\times10^{-10}$~\ecs\ which translates to a luminosity of
$(3-6)\times10^{36}$~\lum.

Finally, \bronb\ was detected with INTEGRAL during a long campaign in
Aug-Sep 2003 at a 18--60 keV flux of $(3.4\pm0.3)\times10^{-11}$~\ecs\
(Revnivtsev et al. 2004; see also Stephen et al. 2005). For a spectrum
like the one measured with Chandra, this extrapolates to a 0.5--10 keV
flux of $5.5\times10^{-11}$~\ecs, which is close to that measured with
Chandra during the same year.  For a power law index as measured
with RXTE (with the same $N_{\rm H}$ as measured with Chandra) the
extrapolated 0.5--10 keV flux is 20\% higher than measured with
Chandra.

In conclusion, the source shows a rather consistent unabsorbed
flux between 8 and $17\times10^{-11}$~\ecs.

\section{Observations of \brona}

The burst from \brona\ was already discussed in detail by Kaptein et
al. (2000); we summarize that discussion here.  The burst onset was on
Sep. 23, 1996, at 07:30:55 UT (MJD~50349.31314). The 2--28 keV photon
flux remained at a plateau for approximately 70 s after that.  The
burst lasted fairly long: after 3.5 min the source set behind the
earth when the 2--28 keV flux had decayed to about 20\% of the peak
value.  The time-resolved spectroscopy revealed clear signs of
photospheric radius expansion which allowed a distance estimate of
$6.5\pm0.5$ kpc for a canonical neutron star if the photosphere is
rich in hydrogen. $N_{\rm H}$ was measured to be
$(3.1\pm1.3)\times10^{22}$~cm$^{-2}$. The unabsorbed bolometric peak
flux was $3.9\times10^{-8}$~\ecs. The energy output is
$>3\times10^{40}$~erg.

Chandra observed \brona\ with ACIS-S in the focal plane on Jul 2,
2004, for a total of 14.8~ksec (net exposure time) starting 06:49:19
UT (obsid 4548).  The CCD frame time was set at the nominal 3.2 s. The
source position, by averaging of the photon positions, was determined
to be $\alpha_{2000.0}=17^{\rm h}18^{\rm m}24\fs144$,
$\delta_{2000.0}=-40^\circ 29\arcmin33\farcs04$ ($l^{II}=-12\fdg72$,
$b^{II}=-1\fdg65$) with a nominal uncertainty of 0\farcs6. This is
2\farcs0 from the ROSAT position which is consistent given that the
ROSAT position has an accuracy of 15\farcs7 (Motch et
al. 1998). Searches in red and infrared plates of the Digital Sky
Survey and in the USNO-B1.0 catalog failed to identify an optical
counterpart brighter than 20 mag. The nearest near-infrared object in
the 2MASS catalog is at 6\farcs3.

\begin{figure}[t]
\includegraphics[height=\columnwidth,angle=270]{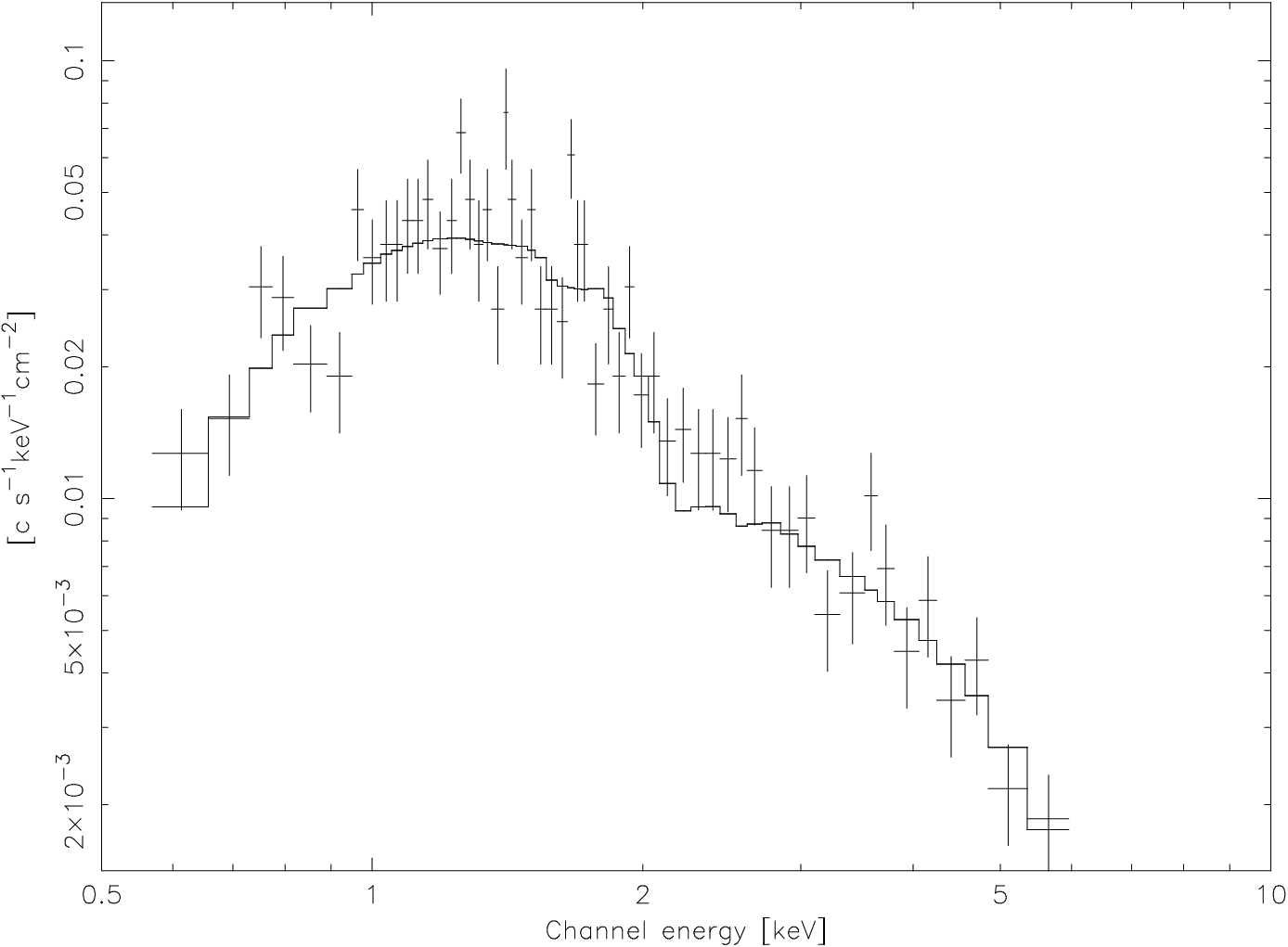}
\caption[]{Chandra/ACIS-S3 trail spectrum of \bronb.
\label{rxj1708sp}}
\end{figure}

\begin{figure}[t]
\includegraphics[height=\columnwidth,angle=270]{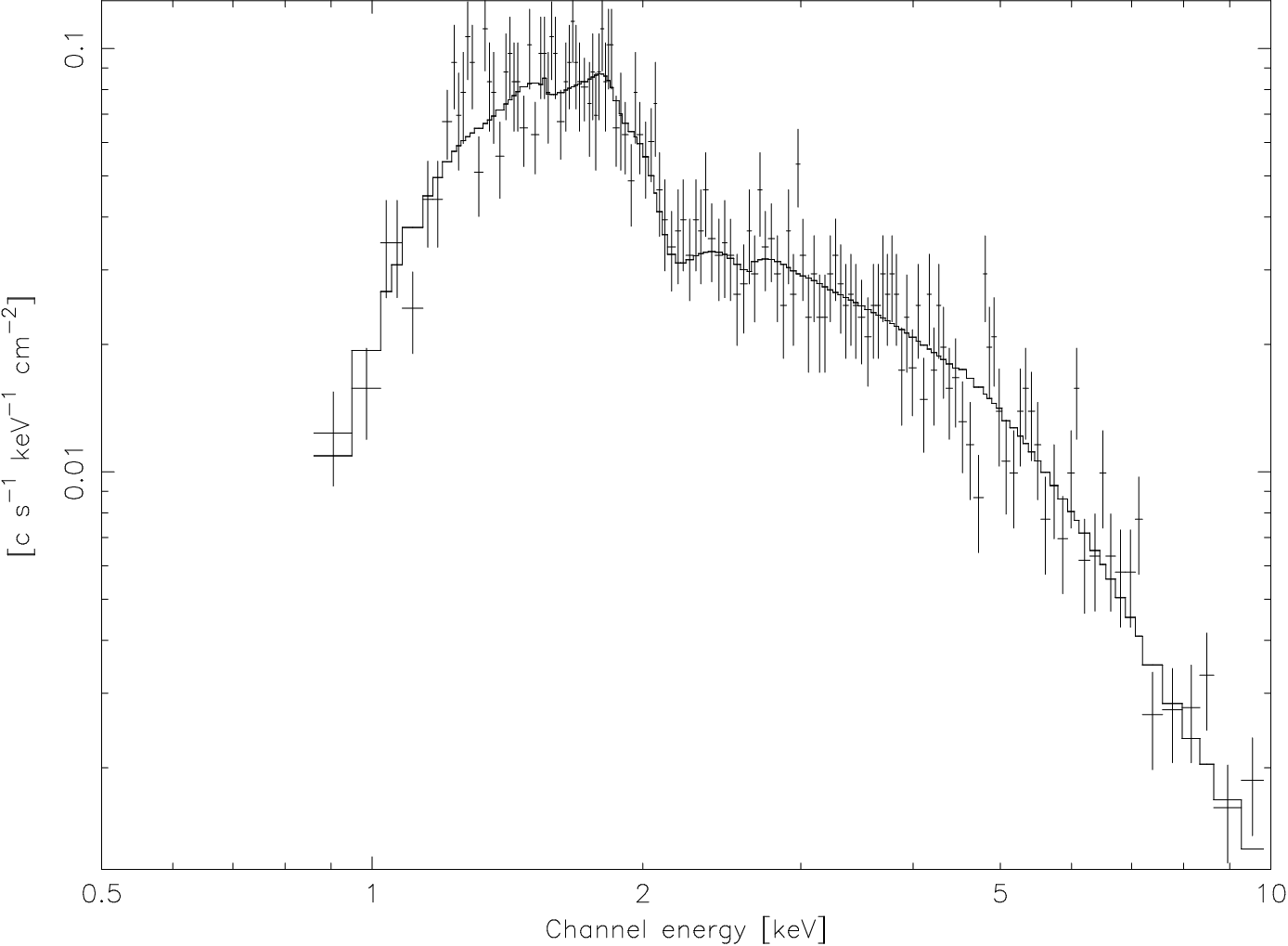}
\caption[]{Chandra/ACIS-S3 spectrum of \brona.
\label{rxj1718sp}}
\end{figure}

The raw source intensity is 0.18 c~s$^{-1}$. This implies significant
pile up for the nominal CCD frame time of 3.2~s, although the
image does not show an obvious drop at the center of the point spread
function (PSF) nor a readout streak. In order to correct for
pile up, the level-1 data were reprocessed to allow inclusion of
'afterglow' events.  From the resulting new level-2 data we extracted
the photons within 2\arcsec\ from the centroid position, while we
determined the background from a nearby circular region with a radius
of 56\arcsec. The source region encompasses 2623 events. The
background is negligible and the source count rate in the full
bandpass is $0.176\pm0.003$~s$^{-1}$. We modeled the data with an
absorbed power law with pile up following Davis et al. (2001). Of
the pile-up model we left free two parameters. One is $\alpha$, the
'grade morphing' parameter (i.e., the probability that $n$ piled-up
photons are not rejected is proportional to $\alpha^{n-1}$). We found
that the fits prefer $\alpha$ between 0.7 and 1, but are not sensitive
to any particular value in that range. The other free pile-up
parameter is the fraction of the PSF subject to pile up. This was
found to be $0.83\pm0.04$. The fit is satisfactory with a goodness of
fit of $\chi^2_\nu=0.94$ for $\nu=135$ degrees of freedom. We find
for the source spectrum $N_{\rm
H}=1.32^{+0.16}_{-0.12}\times10^{22}$~cm$^{-2}$ and a photon index of
$\Gamma=2.09^{+0.22}_{-0.24}$. The spectrum is so much absorbed that
data below 0.8 keV were excluded in the fits. The (extrapolated)
unabsorbed 0.5--10 keV flux before pileup correction is
$(7.1\pm0.1)\times10^{-12}$~\ecs\ and after
$(2.2\pm0.8)\times10^{-11}$~\ecs. The pileup correction is rather
uncertain because the photon flux is very close to the peak in the
curve of the observed versus CCD-incident photon flux (see Fig.~6 in
Davis 2001).  Therefore, we attempted to extract a trail spectrum as
well (see Sect.~\ref{bronbchandra}), despite the apparent absence of
the trail, to obtain a flux unaffected by pile up. We detect a
significant amount of 89 excess photons above a background of
268$\pm4$. This is not sufficient to model the spectral shape
accurately, but it does provide a better handle on the flux than the
piled-up spectrum. Fixing $N_{\rm H}$ and $\Gamma$ to the values found
for the PSF spectrum, we find an unabsorbed 0.5--10 keV flux of
$(9.7\pm1.7)\times10^{-12}$~\ecs. For a distance of 6.5 kpc it
corresponds to $4.9\times10^{34}$~\lum. If the photosphere is
helium-rich, the distance may be up to 9.0~kpc. The implied luminosity
would be $9.4\times10^{34}$~\lum.

We extracted the light curve for photons in the source region. There
are no noticable features in the light curve at a time resolution of
32 to 320~s.

A search through archival data revealed no further detections of
\brona\ except for the ROSAT ones, including ASM data (note that no
other imaging instrument than ROSAT observed this region). There were
two ROSAT observations in March and September of 1994 with the HRI and
the count rates were 0.048$\pm0.008$ and 0.019$\pm0.003$ c~s$^{-1}$
respectively. The number of detected photons was too small for an
accurate spectral analysis. For a spectrum as measured with Chandra,
the count rates translate to unabsorbed 0.5--10 keV fluxes of
$3.0\times10^{-11}$ and $1.2\times10^{-11}$~\ecs\ which is 3.1 and 1.2
times brighter than with Chandra, respectively.

\section{Discussion}

The Chandra measurements show that \bronb\ and \brona\ are actively
accreting. The luminosity levels are clearly above quiescent levels
and the spectra are atypical for quiescence. This confirms the
suspicion that they are persistent rather than classical transient
X-ray sources. We employ as definition for a classical LMXB
transient a source whose flux rises out of a quiescent luminosity of
less than 10$^{33}$~\lum\ and returns to that level within at most a
few months. This excludes so-called 'long-duration' transients like KS
1731-260 and MXB 1658-298 (e.g., Wijnands et al. 2001) that stay on
for up to more than a decade, possibly because the mass transfer rate
from the donor to the accretion disk is temporarily increased. We
regard the latter as temporary persistence, because it is the mass
transfer rate which most likely determines the disk instability
condition.

Most unusual, \brona\ appears to be consistently sub-luminous (i.e.,
below 10$^{36}$~\lum) for a typical persistent source. It is unlikely
that this is due to a smaller-than-usual efficiency in transforming
liberated gravitational energy into radiation, because one would then
expect a burst recurrence time shorter than observed. For a nuclear
energy production between 1.6$\times10^{18}$ and
6$\times10^{18}$~erg~g$^{-1}$ (for pure helium and hydrogen burning,
respectively) it takes an accumulation of 6 to 36$\times10^{21}$~g of
fuel to power a 10$^{40}$~erg burst as observed from \brona\ (Kaptein
et al.  2000). At an accretion rate of
$\approx$6$\times10^{-12}~M_\odot$yr$^{-1}$ as seen with Chandra
this takes at least 0.6 years. For a coverage as with the WFCs
($\sim$70~d net exposure) the chance probability is only 4\% to
observe two or more bursts. If the efficiency at which gravitational
energy is converted into radiation would have been 10 times less, this
probability would increase to 83\% and it would have been likely to
detect multiple bursts.  There are other uncertainties about the
luminosity, such as the extrapolation of 0.5--10 keV to bolometric
luminosity and fluctuations that cannot be observed due to the
insufficient sensitivity of monitoring devices, but the lack of
further burst detections remains a strong indication for a small
accretion level, and we conclude that the low $L$ is truly due to a
smaller $\dot{M}$.

\begin{table*}
\caption{Overview of the burst sources at low persistent emission as
detected with the Wide Field Cameras on board BeppoSAX. The distances
were determined by assuming burst peak fluxes to be equal (for
radius-expansion bursts) to or smaller than the Eddington limit of a
canonical neutron star, and have a typical uncertainty of 30\%.
$\tau$ is the overall e-folding decay time of the burst.
\label{tab1}}
\begin{center}
\begin{tabular}{lccccccc}
\hline
Object & $l^{II}$ & $b^{II}$ & Distance & $L_{\rm pers}$ & Follow up instr.&$\tau$ & Ref.$^\ddag$\\
     &      &  & (kpc) & (10$^{32}$~\lum) & (delay in yr) & (s) \\
\hline
SAX\,J1324.5--6313        & $306\fdg6$ & $-0\fdg6$  & $<6.2$ &  $<4$ & Chandra (+4.1)     &  6.0 & 1,2  \\
\bronb                    & $352\fdg8$ & $+4\fdg7$  & 13   & 15000 & Chandra (+3.4)     & 300  & 3 \\
\brona                    & $347\fdg3$ & $-1\fdg7$  & 8    &  700  & Chandra (+7.8)     & 47.5 & 3,4 \\
SAX\,J1752.4--3138        & $358\fdg4$ & $-2\fdg6$  & 9.2  & $<3$  & Chandra (+2.0)     & 21.9 & 2,5 \\
SAX\,J1753.5--2349        & $5\fdg3$   & $+1\fdg1$  & $<8$ & $<4$  & Chandra (+5.1)     & 8.9  & 2,6 \\
SAX\,J1818.7+1424         & $42\fdg3$  & $+13\fdg7$ & $<9.4$ & $<4$  & Chandra (+3.9)     & 4.5  & 1,2 \\
SAX\,J2224.9+5421         & $102\fdg6$ & $-2\fdg6$  & $<7.1$  &  $<12$   & BeppoSAX (+0.001)  & 2.6  & 1 \\
\hline
\multicolumn{7}{l}{$^\ddag$ 
1. Cornelisse et al. 2002a, 2. Cornelisse et al. 2002b, 3. this paper,
4. Kaptein et al. (2000), 5. Cocchi et al.}\\
\multicolumn{7}{l}{(2001), 6. in 't Zand et al. (1998)}\\
\end{tabular}
\end{center}
\end{table*}

\bronb\ appears to be more luminous by an order of magnitude, but
still less luminous than most persistent LMXBs, hovering around the
10$^{36}$~\lum\ mark. It may be regarded as the far-away equivalent of
systems like 2S 0918--549 (e.g., Jonker et al. 2001 and Cornelisse et
al. 2002a), 4U 0614+091 (Piraino et al. 1999) and various other
faint LMXBs.

The persistent nature at such a low accretion level suggests the
possibility of a relatively small binary orbit.  Smaller orbits generally
harbor smaller accretion disks which are easier to
photo-ionize completely through X-ray irradiation from the central
source and, thus, can easier
keep the accretion onto the neutron star going at lower levels (e.g.,
van Paradijs 1996). For a donor of solar composition, the critical
mass accretion rate $\dot{M}_{\rm crit}$ below which the disk becomes
unstable was by Dubus et al. (1999) derived to be
\begin{eqnarray*}
\dot{M}_{\rm crit} & = & 6.6\times10^{-22}R_{\rm rim}^{2.1} C^{-0.5}
\hspace{5mm} M_\odot~{\rm yr}^{-1} 
\end{eqnarray*}
where $R_{\rm rim}$ is the radius of the outer disk edge in km and 
$C$ a somewhat uncertain factor $C\approx1$; a neutron star mass of
1.4~$M_\odot$ has been assumed. The actual accretion rate needs to be
higher than this for the source to be persistent.  The
mass accretion rate for \brona\ was the lowest of our two sources,
between $\dot{M}_{\rm
obs}=4.1\times10^{-12}$ and $2.4\times10^{-11}$ $M_\odot~$yr$^{-1}$
(assuming that all liberated gravitional energy is transformed to
radiation).  This implies $R_{\rm rim}<1.0\times10^5$~km. 
The uncertainty in this number is considerable.
A $C$ value between $-50$ to $+50$\% from the nominal value gives
rise to an uncertainty in $R_{\rm rim}$ of $\pm$20\%. 

\begin{table}
\caption[]{List of sources previously considered as low-$L$ bursters but now
established as transients\label{tab2}}
\begin{center}
\begin{tabular}{lll}
\hline
Object & $L_{\rm pers}$ & Ref.$^\ddag$ \\
       & (10$^{32}$~\lum) & \\
\hline
GRS\,1741.9--2853  & $2\times10^5$ & 1 \\
SAX\,J1828.5--1037 & 340 (0.02 yr after burst) & 2 \\
SAX\,J1806.5--2215 & $3\times10^4$ & 3 \\
\hline
\end{tabular}

\noindent
$^\ddag$ 1. Muno, Baganoff \& Arabadjis 2003, 2. Hands et al. 2004,
3. Cornelisse et al. 2002a
\end{center}
\end{table}

Given the persistence, the accretion rate may be a good representation
of the mass transfer rate from the donor.  One then wonders whether it
is possible to 'construct' a binary with the implied transfer rate and
accretion disk size. Such a low mass transfer rate implies a
degenerate donor star. If the donor is a white dwarf and the accretor
a 1.4~$M_\odot$ neutron star, the mass transfer rate fixes directly
the basic binary parameters through the mass-radius relation of the
white dwarf\footnote{the white dwarf mass-radius relation is
parameterized by $R/R_\odot=0.0115(M_\odot/M_{\rm WD})^{1/3}$ which
compares well to the helium white dwarf models in Deloye \& Bildsten
2003. Note that we need a helium white dwarf in order to fuel the
X-ray burst.}, the requirement that the donor fills its Roche lobe,
gravitational radiation driving the mass transfer and Kepler's third
law. The donor mass then is given by $M_{\rm WD} = 3.28
\dot{M}^{3/14}~M_\odot$ with $\dot{M}$ in $M_\odot$yr$^{-1}$. For
$\dot{M}$ between $4.1\times10^{-12}$ and
$2.4\times10^{-11}$~$M_\odot$yr$^{-1}$, this implies $M_2 =
0.012-0.017~M_\odot$, $P_{\rm orb}=0.6-0.9$~hr, and a neutron star
Roche lobe average radius of $(2.3-3.0)\times10^5$~km. Therefore, the
Roche lobe size is at least two times larger than the part of the
accretion disk that can be ionized by irradiation from the central
source and the picture for \brona\ is not completely
consistent. Perhaps the 3 flux measurements with ROSAT and Chandra are
not representative for the source. We hope to find out by carrying out
more flux measurements with Chandra in the future.

Our observations show that the two low-$L$ bursters investigated here
are clearly different from the four other cases studied with Chandra
by Cornelisse et al. (2002b) as well as SAX~J2224.9+5421 (Cornelisse
2002a), see Table~\ref{tab1}. Those were {\em never} seen in an
actively accreting state, suggesting that they are {\em transient}
instead of {\em persistent} low-$L$ sources. It is also interesting to
note that the X-ray bursts in those other cases are rather short
compared to the $>3.5$~min for \brona\ and 10~min for \bronb. All are
shorter than 1~min, with the most extreme case being SAX~J2224.9+5421
which lasted less than 10~s.  The persistent character of our two
systems may be consistent with the long duration of the bursts. If the
accretion on the neutron star remains low, like in the persistent
sources, instead of briefly becoming high during fast accretion events
in transients, ignition conditions may take longer to develop and
flashes to be more energetic and, thus, longer.

It is difficult to obtain independent evidence for a short
orbital period explanation. The X-ray light curves show no evidence
for orbital modulation. Perhaps the easiest way to confirm this is to
search for the optical or infrared counterparts and determine the
optical to X-ray flux ratio, which is presumed to be a good indicator
for the orbital period (van Paradijs \& McClintock 1994).  The limits
obtained from the Digitized Sky Survey are not constraining to make a
distinction between short and long (cf, Wachter et al. 2005a and
2005b). A deeper search is necessary. The excellent Chandra positions
will be instrumental in this.

\acknowledgement 

We are grateful to Ron Remillard for analyzing RXTE ASM data on both sources
discussed in this paper. Mike Revnivtsev, Peter Jonker, Elisa
Constantini, Rudy Wijnands and Frank Verbunt are thanked for useful
discussions, and the anonymous referee for important suggestions on
the interpretation. JZ acknowledges support
from the Netherlands Organization for Scientific Research (NWO). This
research has made use of the SIMBAD database, operated at CDS,
Strasbourg (France), and of NASA's Astrophysics Data System.

\end{document}